\documentclass{article}

\usepackage{graphicx}

\begin{document}

\title{Passage of a Bessel Beam Through a Classically Forbidden Region}
\author{D. Mugnai \\
{\small\it Istituto di Ricerca sulle Onde Elettromagnetiche ``Nello Carrara''
- CNR,} \\
{\small\it Via Panciatichi 64, 50127 Firenze, Italy} }

\date{}
\maketitle

\section{Introduction}

The motion of an electromagnetic
wave, through a classically-forbidden region, has recently attracted  renewed interest  because of
its implication with  regard to the theoretical and experimental
problems of superluminality.
From an experimental point of view, many papers
provide an evidence of
superluminality in different physical systems\cite{guide,nimtz,diffra,prism,chiao1,franc}.
Theoretically, the problem of a passage through a forbidden gap
has been treated by considering plane waves at oblique
incidence into a plane parallel layer of a medium with a refractive
index smaller than the index of the surrounding medium, and also
confined (Gaussian) beams, still at oblique
incidence\cite{bos,chiao2,atti,io}.
In the present paper the case of  a Bessel beam is examined,
at normal incidence into the layer (Secs. II and III), in the scalar approximation (Sec. IV) and by developing also a vectorial treatment (Sec. V).
Conclusions are reported in Sic. VI.

\section{The Bessel beam}

An interesting solution of the wave equation is represented by the
Bessel beam with axial symmetry having, as known\cite{dur,saa}, 
the following expression:

\begin{equation} 
u(\rho ,\:\psi ,\:z)=AJ_0(k_0n\rho\sin\theta_0) \exp(ik_0nz\cos \theta_0)\:,
\label{u}
\end{equation}
where $A$ is an amplitude factor, $J_0$  denotes the
zero-order Bessel function of first kind,  $\rho ,\:\psi ,\:z$
are cylindrical coordinates (Fig. 1a), $\theta_0$ is the parameter
of the beam (Axicon angle),
$n$  is the refractive index of the medium where the
beam propagates, and $k_0$ is the wavenumber in the vacuum. The beam
is independent of the angular coordinate $\psi$. 
The time factor $\exp(-i\omega t)$ is omitted in Eq. (\ref{u}).

The unusual features of a Bessel beam are that its phase propagates
(in the $z$ direction) with a velocity $v=c/(n\cos\theta_0)$  larger
than $c/n$\cite{saa,noiprl2}, and that it does not changes its shape during propagation (the amplitude is independent of $z$).
The situation is similar to what 
occurs  when only two plane waves interfere, the only difference
being that the two-wave interference pattern occupies the whole space,
while the field (\ref{u}) is practically limited to a restricted
zone of space $(k_0n \rho\sin\theta_0 < 2.4,\:2.4$  being the first
zero of $J_0$). 
In this connection, it is worth noting that the field
of Eq. (\ref{u}) is not properly a beam, since it is not limited by
a caustic surface, inasmuch as $J_0$ oscillates when its argument
tends to infinity.

\begin{figure}
\begin{center}
\includegraphics[width=.5\textwidth]{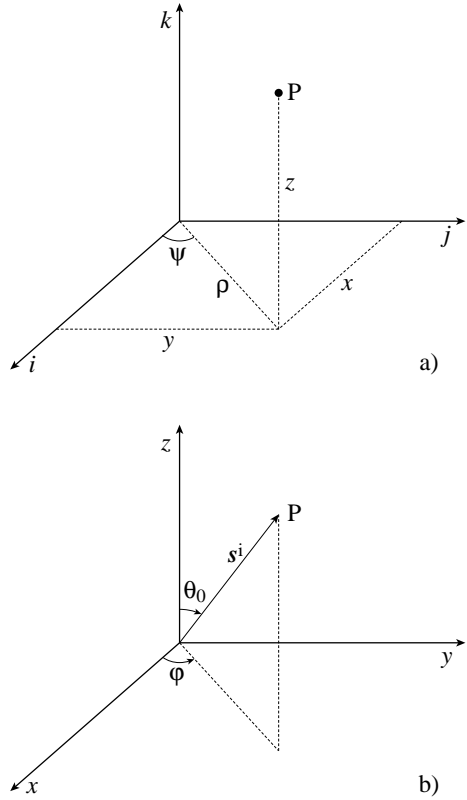}
\end{center}
\caption{The coordinate systems}
\end{figure}

The meaning of the parameter 
$\theta_0$ is the following.
Let us refer the space to a system of Cartesian coordinates
$x,\: y,\: z$, (unit vectors {\bf i, j, k}), such that

\begin{eqnarray} 
x&=& \rho\cos\psi \nonumber \\
y&=& \rho\sin\psi \nonumber \\
z&=&z
\label{cil}
\end{eqnarray}
Let us also consider a system of spherical coordinates
$r,\:\theta,\:\varphi$, with origin in the origin of the
Cartesian coordinates, and the $\theta = 0$ semiaxis coinciding
with the positive $z$-axis.

Let us now consider a set of plane waves, with the same amplitude $A d\varphi$,
and with directions of propagation {\bf s} = $\alpha${\bf i}+
$\beta${\bf j} + $\gamma${\bf k} making the same angle $\theta_0$ with the
$z$-axis. Each wave can be written as

\begin{equation} 
A\exp [ik_0n(\alpha x+\beta y+\gamma z)] \,d\varphi \:,
\label{onda}
\end{equation}		
where the well-known relations hold (see Fig. 1b):

\begin{equation}  
\alpha =\sin\theta_0\cos\varphi\,,\:\:\:\beta =\sin\theta_0\sin\varphi\,,
\:\:\: \gamma =\cos\theta_0 \,.
\end{equation}
If we integrate expression (\ref{onda}) over $d\varphi$ between 0
and 2$\pi$, and recall the properties of the Bessel function $J_0$\cite{wat},
we arrive at Eq. (\ref{u})

\begin{eqnarray}  
&A&\int_0^{2\pi}\,\exp [ik_0n (\alpha x+\beta y+\gamma z)]\,d\varphi \nonumber \\
=&A& \,\exp(ik_0nz\cos\theta_0)
\int_0^{2\pi}\,\exp [ik_0n\rho (\cos\varphi\cos\psi +\sin\varphi\sin\psi )
\sin\theta_0 ]d\varphi  \nonumber \\
=&A& \,2\pi J_0(k_0n\rho\sin\theta_0 )\exp (ik_0nz\cos\theta_0 )
\label{bes}
\end{eqnarray}
If beam (\ref{u}) impinges normally into a plane parallel layer
of refractive index $n^\prime$, limited by the planes $z=0$ and
$z=d$ (Fig.2), the plane waves (\ref{onda}) form an incidence
angle $i=\theta_0$ which may be larger than the limit angle
$i_0= \arcsin (n/n^\prime )$. 
In this case,  the plane waves undergo
total reflection, and the layer is a classically-forbidden region
for the Bessel beam, in spite of the fact that the beam impinges
normally into the layer.

\begin{figure}[t]
\begin{center}
\includegraphics[width=.5\textwidth]{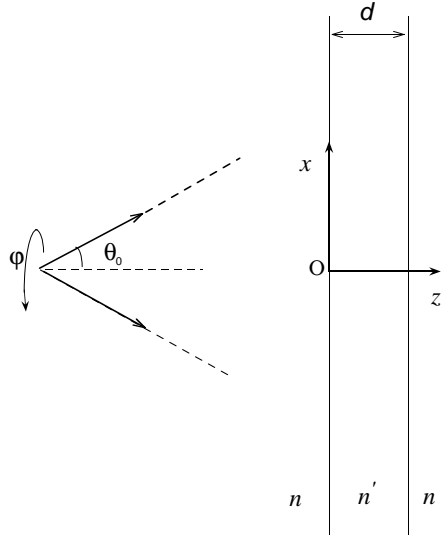}
\end{center}
\caption{The layer and the impinging Bessel 
beam characterised by the Axicon angle $\theta_0$.}
\end{figure}

\section{Total reflection for a Bessel beam}

The preceding expansion of a Bessel beam in plane waves whose
directions of propagation cover a conical surface of semiaperture
$\theta_0$ immediately indicates what happens when the Bessel beam
impinges (at $z = 0$) into a plane surface separating two media of
different refractive indexes $n$ and $n^\prime$. Each plane wave
forms the same incidence angle $\theta_0$  and, hence, also the same
refraction angle $\theta^\prime$, satisfying

\begin{equation}  
n^\prime\sin\theta^\prime = n\sin\theta_0 \,.
\end{equation}
The refracted waves have the same amplitude, therefore, their
superposition $u$ is a Bessel beam with parameter $\theta^\prime$:
  
\begin{equation}  
u(\rho ,\psi ,z)=A^\prime J_0(k_0n^\prime\rho\sin\theta^\prime )
\exp (ik_0n^\prime z\cos\theta^\prime )
\end{equation}
Here, it may be interesting to note that, if $\cos\theta^\prime$
is real, the phase of the Bessel beam for $z > 0$ propagates in the
$z$-direction, normally therefore to the boundary. 
The phase velocity
turns out to be $c/(n^\prime \cos\theta^\prime )$,
i.e. larger than the light velocity in the 
$n^\prime$ medium. 
If $\cos\theta^\prime$ is purely imaginary (that is,
for $\theta_0$ larger than $i_0$), the single
plane waves composing the incident Bessel beam are in total reflection,
and therefore give rise to plane refracted waves the phase of which
propagates parallel to the boundary. However, for the refracted
Bessel beam, there is no phase propagation for $z > 0$: there is a
sort of stationary field.
Here as follows, only the latter case will be considered, namely:

\begin{equation} 
\sin\theta^\prime =\,\frac{n}{n^\prime}\,\sin\theta_0 >1
\label{sint}
\end{equation}
with the notation
\begin{equation}  
\cos\theta^\prime =i\Gamma, \:\:\:\:\:\:\:\:\:  (\Gamma >0)\,.
\end{equation}

\section{The tunneling effect in the scalar approximation}

If the beam of Eq. (\ref{u}) impinges normally into the layer
of Fig. 2, it gives rise: \newline
{\it i} - on the left of the first boundary ($z < 0$), to a reflected
Bessel beam which propagates in the negative direction of the z-axis;
\newline
{\it ii} - inside the layer ($0 < z < d$), to two Bessel beams, a ``progressive" one and a 
``regressive" one; \newline
{\it iii} - on the right of the second boundary ($z > d$),  to one
transmitted Bessel beam, which propagates in the positive $z$-direction.

The continuity conditions of 
the total field at the boundaries $z = 0$ and $z  = d$ may be 
easily satisfied with a suitable choice of the complex amplitudes 
of the beams, and of the parameter $\theta^\prime$ in the argument 
of the Bessel beams inside the layer.
Using standard procedure, let us write the incident field in the form

\begin{equation} 
u^i(\rho ,\psi ,z)=A^i\,2\pi J_0(k_0n\rho\sin\theta_0 )
\exp (ik_0n z\cos\theta_0 ) \:\:\:\:\: (z\leq 0)
\label{ui}
\end{equation} 
the reflected field in the form

\begin{equation} 
u^r(\rho ,\psi ,z)=A^r \,2\pi J_0(k_0 n \rho \sin\theta_0 )
\exp (-ik_0n z\cos\theta_0) \:\:\:\:\:\:\: (z\leq 0)
\label{ur}
\end{equation}		
the transmitted field in the form

\begin{equation} 
u^t(\rho ,\psi ,z)=A^t\, 2\pi J_0(k_0 n \rho \sin\theta_0)
\exp (i k_0 n (z-d) \cos\theta_0 ) \:\:\:\:\: (z\geq d)
\label{ut}
\end{equation} 		
and, lastly, the progressive and regressive fields, respectively, in the forms

\begin{eqnarray}   
u^+(\rho ,\psi ,z) &=& A^+\,2\pi J_0(k_0 n^\prime \rho \sin\theta^\prime )
\exp (-k_0 n^\prime z \Gamma )  \nonumber \\
u^-(\rho ,\psi ,z) &=& A^-\,2\pi J_0(k_0 n^\prime \rho \sin\theta^\prime )
\exp (k_0 n^\prime z \Gamma) \, \:\:\:\:\:\:\:\:\:  (0\leq z \leq d).
\label{uin}
\end{eqnarray}
Due to the exponential dependence of $u^+$  and $u^-$ on $z$, the
progressive and regressive beams inside the layer may be denoted
as ``evanescent" Bessel beams.
     
The ratio $R = A^r/A^i$  is the reflection coefficient of the 
layer for the Bessel beam; the ratio $T = A^t/A^i$ is 
the transmission coefficient.

By denoting the total field at any point of the
space by $u^{tot}$, the boundary conditions can be written as

\begin{eqnarray}   
u^{tot}(z=0-) = u^{tot}(z=0+) \nonumber \\
u^{tot}(z=d-0) = u^{tot}(z=d+0) 
\label{utot}
\end{eqnarray}		
and

\begin{eqnarray}  
\frac{\partial}{\partial z}\,  u^{tot}(z=0-) = \frac{\partial}
{\partial z}\, u^{tot}(z=0+) \nonumber \\
\frac{\partial}{\partial z}\,  u^{tot}(z=d-0) = \frac{\partial}
{\partial z}\,  u^{tot}(z=d+0)
\label{dutot}
\end{eqnarray}
		
Upon the introduction of Eqs. (\ref{ui}) to (\ref{uin}) into Eqs.
(\ref{utot}) and (\ref{dutot}), the following conditions are found:

\begin{eqnarray}  
A^i+A^r &=& A^+ +A^- \nonumber \\
(A^i -A^r) n \cos\theta_0  &=&  i n^\prime \,\Gamma ( A^+ -A^- )\nonumber \\
A^+ e_1+A^- e_2 &=& A^t \nonumber \\
i n^\prime \,\Gamma ( A^+ e_1 -A^- e_2 ) &=& A^t n \cos\theta_0 \,,
\label{cont}
\end{eqnarray}
where $e_1=\exp(-k_0 n^\prime d \Gamma ),\: e_2=1/e_1$.

The solution of system (\ref{cont}) is easily 
found and we have\footnote{The evaluation of the reflection coefficient, which maybe derived from Eq. (\ref{cont}), is of no interest in the present paper.}

\begin{eqnarray}  
\frac{A^+}{A^i}\,=\, \frac{e_2}{2 n^\prime \,\Gamma}
(n^\prime \Gamma -in\cos\theta_0) T  \nonumber \\
\frac{A^-}{A^i}\,=\, \frac{e_1}{2 n^\prime \,\Gamma}
(n^\prime \Gamma + in\cos\theta_0) T  \nonumber \\
\frac{A^t}{A^i}\,= T=\, \frac{4inn^\prime\Gamma\cos\theta_0}
{e_2(n\cos\theta_0+in^\prime \Gamma )^2-e_1(n\cos\theta_0 - in^\prime 
\Gamma )^2}    
\label{asua}
\end{eqnarray}
From the point of view of the field iside to the forbidden
region, it should be noted that $A^+$ and $A^-$ are complex
quantities that depend on $k_0$
and on the geometric characteristics of the system.
If we denote the argument of $T$ by $\Phi_T$, and
if we introduce a quantity $\Phi^\prime$ such that

\begin{equation} 
\tan\Phi^\prime =\,\frac{n\cos\theta_0}{n^\prime\Gamma}
\end{equation}				
we can write

\begin{eqnarray}  
A^+ = |A^+| \exp [i(\Phi_T - \Phi^\prime )] \nonumber \\
A^- = |A^-| \exp [i(\Phi_T + \Phi^\prime )]\,.			
\end{eqnarray}
Accordingly, the internal total field can be written as

\begin{eqnarray}   
u^{tot}&=&u^+ + u^- = 2\pi\,\exp (i\Phi_T)J_0 (k_0 n\rho\sin\theta_0) \times 
\nonumber \\
& \times &\left[|A^+|\exp (-k_0n^\prime z\Gamma ) \exp (-i\Phi^\prime )
+|A^-|\exp (k_0n^\prime z\Gamma ) \exp (i\Phi^\prime ) \right] \,,
\end{eqnarray}	
which shows that, inside the forbidden layer, the total field has a phase 
$\Phi = \Phi_T +\eta (z)$, with $\eta (z)$ such that

\begin{eqnarray} 
\tan\eta (z) &=& \,\frac{|A^-|\exp (k_0n^\prime z\Gamma ) -
|A^+|\exp (-k_0n^\prime z\Gamma )}
{|A^-|\exp (k_0n^\prime z\Gamma ) +|A^+|\exp (-k_0n^\prime z\Gamma )}
\tan\Phi^\prime
\nonumber \\
&=& -\tanh [ k_0n^\prime \Gamma (d-z)] \tan \Phi^\prime \,,
\label{fiz}
\end{eqnarray}	
which propagates in the direction of the positive $z$-axis. 

Equation (\ref{fiz}) may be used to evaluate $|{\rm grad}\eta | =
\partial\eta /
\partial z$, the wavelength $\lambda^\prime = \lambda^\prime (z)$
and the phase velocity $v^\prime = v^\prime (z)$:

\begin{eqnarray}  
|{\rm grad} \eta |&=& \frac{ k_0n\cos\theta_0\,\cos^2[\eta (z)]}{\cosh^2
[k_0 n^\prime \Gamma (d-z)]} \nonumber \\
\lambda^\prime &=& \frac{2\pi}{|{\rm grad} \eta |}= \frac{\lambda_0}
{n \cos\theta_0} \,\frac{\cosh^2 [k_0 n^\prime \Gamma (d-z)]}
{\cos^2[\eta (z)]}  \nonumber \\
v^\prime &=& \frac{\omega}{|{\rm grad} \eta |}= \frac{c}{n \cos\theta_0}
\,\frac{\cosh^2 [k_0 n^\prime \Gamma (d-z)]}
{\cos^2[\eta (z)]} \,,
\label{grad}
\end{eqnarray}
where $\lambda_0 = 2\pi /k_0$ denotes the free space plane-wave
wavelength. It is interesting to note that for $z\rightarrow d,\: v^\prime\rightarrow c/(n\cos\theta_0)$,
which is equal to the phase velocity of the trasmitted ($z>d$)
and incident ($z<0$) fields. This is a phenomenon similar
to the one reported in Ref.\cite{io}.

The phase difference $\Delta\Phi$ of the total internal field at
$z = d$ from $z = 0$ is given by

\begin{equation}  
\Delta\Phi  = \arctan [\tanh (k_0 n^\prime \Gamma d)
\tan \Phi^\prime ] \,,
\label{dfi}
\end{equation}
while, as to the transmitted field, it can be noted (see Eq. (\ref{ut}))
that its phase at $z = d$  is

\begin{equation}
\Phi_T= \arctan \left[ \frac{n^2 \cos^2\theta_0 - {n^\prime}^2\Gamma^2}
{2n  n^\prime \Gamma \cos\theta_0} \tanh (k_0n^\prime\Gamma d)\right]\,.
\label{ft}
\end{equation}
Thus, $\Phi_T$  is equal to the phase shift of the transmitted
beam at $z = d$ with respect to that of the incident field at $z = 0$.
By  comparing  Eqs. (\ref{dfi}) and (\ref{ft}), it appears that
 $\Phi_T$ does not coincide with the phase gained by the internal field in
passing from $z = 0$ to $z = d$.

\section{The vectorial treatment}

In this Section we use the vectorial algorithm for analysing  the propagation of Bessel beams.

\subsection{Quasi-TE and quasi-TM beams}

For a vectorial treatment of the propagation of a Bessel beam it is sufficient to consider, for example, the function of Eq. (\ref{u}) as the tangential component of the electric field {\bf E}.
Then, the Maxwell equations allow us to determine the longitudinal component of {\bf E}, and the magnetic field {\bf H} as well. 
The longitudinal component of {\bf E} turns out to vanish on the $z$-axis, for $\rho =0$, where the tangential component has its maximum.
Thus, such field ({\bf E}, {\bf H})  may be named {\it quasi}-TE.
If the field ({\bf E}, {\bf H}) impinges normally on the layer of Fig. 2, it gives rise, as in the scalar approximation, to a reflected Bessel beam for $z<0$, to a transmitted beam for $z>d$, and to a progressive and a regressive Bessel beams for $0<z<d$.
The progressive and regressive beams are evanescent if Eq. (\ref{sint}) holds.

Analogously, we can consider the function $u$ of Eq. (\ref{u}) as the tangential component of a magnetic field {\bf H}', then from the Maxwell equations we can derive the longitudinal component of {\bf H}' and the associated electric field {\bf E}'. 
The field  ({\bf E}', {\bf H}') may be named {\it quasi}-TM.

The complex amplitudes of all the above beams may be determined by imposing,
at the two boundaries, the continuity conditions for the tangential component of both the total electric and magnetic fields.
This allows to determine in both cases the total field inside the forbidden region, its wavelength and its phase velocity.

Since the treatment is a little cumbersome, we report that different wavelength and different phase velocity are found for the quasi-TE and quasi-TM cases;
here we limit ourselves to develop the simpler analysis in the TE and TM cases.
To this end, let us consider a plane
TE  and a plane TM waves with a direction of propagation {\bf s}$^i$ =
$\alpha${\bf i} + $\beta${\bf j}
+ $\gamma${\bf k} = $\sin\theta_0\cos\varphi${\bf i} +
$\sin\theta_0\sin\varphi${\bf j} + $\cos\theta_0${\bf k}.
Let them reflect and refract through the first boundary,
then  reflect and refract at the second boundary.
Lastly, 
we integrate the three Cartesian components with respect to
$\varphi$ between 0 and $2\pi$.

\subsection{The TE case}

This case has already been treated in Ref.\cite{atti,io}, but in the
particular case of $\varphi = 0$. Here, we have to
generalise the results obtained therein.

For the incident field {\bf E}$^i = E_x${\bf i} $+E_y${\bf j} 
$(E_z = 0)$, let us put
\begin{eqnarray}  
E_x&=&A_x\exp [\,ik_0n(x\sin\theta_0\cos\varphi +y\sin\theta_0\sin\varphi
+z\cos\theta_0 )\,] d\varphi \nonumber \\
E_y &=&A_y\exp [\,ik_0n(x\sin\theta_0\cos\varphi +y\sin\theta_0\sin\varphi
+z\cos\theta_0 )\,] d\varphi \,.
\label{exey}
\end{eqnarray}
Since {\bf E}$^i$ is normal to {\bf s}$^i$, the following
relation must hold:
	
\begin{eqnarray*}
A_x\alpha +A_y\beta =0 \:.
\end{eqnarray*}
Hence,

\begin{equation} 
A_x\cos\varphi + A_y \sin\varphi =0   \,.
\label{axay}
\end{equation}
Equation (\ref{axay}) shows that $A_x$ and $A_y$ depend on $\varphi$. 
The solution of
Eq. (\ref{axay}), which remains finite for any value of $\varphi$, is

\begin{eqnarray}  
A_x &=& A \sin\varphi \nonumber \\
A_y &=& -A\cos\varphi      \,,
\label{axay1}
\end{eqnarray}
where $A$ is a constant.

At this point, insertion of Eqs. (\ref{axay1}) into Eqs.
(\ref{exey}) and integration with respect to $\varphi$ yields
(by putting $a=2\pi A$)

\begin{eqnarray}  
E_{x, beam} &=& i\,a\sin\psi J_1 (k_0 n \rho\sin\theta_0)
\exp \left( i k_0 n z\cos\theta_0 \right)  \nonumber \\
E_{y, beam} &=& - i\,a\cos\psi J_1 (k_0 n \rho\sin\theta_0)
\exp \left( i k_0 n z\cos\theta_0 \right) \,,
\label{e}
\end{eqnarray}
where $J_1$ is the Bessel function of the first order.

\begin{figure}[t]
\begin{center}
\includegraphics[width=.6\textwidth]{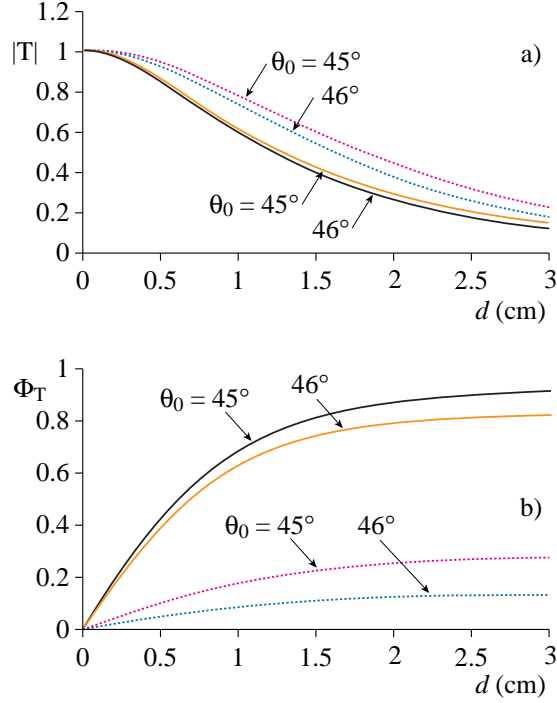}
\end{center}
\caption{Amplitude (a) and phase (b) of the transmission 
coefficients $T$ in the TE case (solid lines) and in the TM case 
(dotted lines) plotted vs $d$  for two values of the parameter 
$\theta_0$. Other
parameter values are: $\omega =60,\:n=1.5,\:n^\prime =1$.}
\end{figure}

The magnetic field {\bf H}$^i = H_x${\bf i} + $H_y${\bf j}
+ $H_z${\bf k} related to the {\bf E}$^i$ field  of the plane
wave is found to be

\begin{eqnarray}  
H_x &=& \frac{A}{Z}\, \cos\theta_0\cos\varphi
\exp [\,ik_0n (x\sin\theta_0\cos\varphi +y\sin\theta_0\sin\varphi
+z\cos\theta_0 )\,] d\varphi \nonumber \\
H_y &=& \frac{A}{Z} \cos\theta_0\sin\varphi
\exp [\,ik_0n (x\sin\theta_0\cos\varphi +y\sin\theta_0\sin\varphi
+z\cos\theta_0 )\,] d\varphi \nonumber \\
H_z &=& -\,\frac{A}{Z} \sin\theta_0
\exp [\,ik_0n (x\sin\theta_0\cos\varphi +y\sin\theta_0\sin\varphi
+z\cos\theta_0 )\,] d\varphi \,,
\end{eqnarray}
where $Z=Z_0/n$, and $Z_0$ is the free space impedance.
Hence, upon integration over $\varphi$, we obtain

\begin{eqnarray} 
H_{x, beam} &=& \frac{i\,a}{Z}\cos\theta_0\cos\psi J_1 (k_0 n \rho\sin\theta_0)
\exp \left( i k_0 n z\cos\theta_0  \right)  \nonumber \\
H_{y, beam} &=&  \frac{i\,a}{Z}\cos\theta_0\sin\psi J_1 (k_0 n \rho\sin\theta_0)
\exp \left( i k_0 n z\cos\theta_0  \right) \nonumber \\
H_{z, beam} &=& - \,\frac{a}{Z}\sin\theta_0 J_0 (k_0 n \rho\sin\theta_0)
\exp \left( i k_0 n z\cos\theta_0 \right) \,.
\label{h}
\end{eqnarray}
A field described by Eqs. (\ref{e}) and (\ref{h}) can be
denoted as a Bessel beam (of the first order), and has
properties very similar to those of the Bessel beam of
Eq. (\ref{u}). However, only the longitudinal component of the
magnetic field is of the type of Eq. (\ref{u}), that is, picked
on the axis at
$\rho = 0$; all other components for $\rho = 0$ vanish. 
A TE plane wave like the one described above gives rise, at the
incidence on the layer of Fig. 2, to a reflected first-order
Bessel beam and to two transmitted ``evanescent" Bessel beams.
One of these is progressive and the other regressive, with coefficients $A^+$ and
$A^-$ given by the first two Eqs. (\ref{asua}). 
At the
second boundary of the layer, a transmitted first-order Bessel
beam originates, with amplitude $A^t$ given by the third
Eq. (\ref{asua}).
Accordingly, the wavelength $\lambda_{TE}$ and the phase
velocity $v_{TE}$  inside the layer are again given
by Eqs. (\ref{grad}).

\subsection{The TM case}

The TM case may be treated in a way analogous to that of the TE case,
arriving at similar results. The only difference lies 
in the fact that $n$ must be replaced by
$n^\prime$, and vice versa, in the coefficients of system
(\ref{cont}) (not in the propagation factors $e_1$ and $e_2$),
analogously to what happens in the Fresnel formulas for the
reflection and transmission coefficients of a real plane wave
at a plane interface. Accordingly, the expression of
$A^+,\: A^-,\: A^t$  and $\Phi^\prime$  must be replaced by

\begin{eqnarray}  
\left(\frac{A^+}{A^i}\right)_{TM}\, &=& \, \frac{e_2}{2 n \Gamma}
(n \Gamma -in^\prime \cos\theta_0)  \nonumber \\
\left(\frac{A^-}{A^i}\right)_{TM}\, &=& \, \frac{e_1}{2 n \Gamma}
(n \Gamma + in^\prime \cos\theta_0)   \nonumber \\
\left(\frac{A^t}{A^i}\right)_{TM}\, &=& T_{TM} = \, \frac{4inn^\prime\Gamma\cos\theta_0}
{e_2(n^\prime \cos\theta_0+in\Gamma )^2-e_1(n^\prime \cos\theta_0 - 
in\Gamma )^2}   \nonumber \\ 
\tan\Phi^{\prime\prime} &=& \,\frac{n^\prime\cos\theta_0}{n\Gamma} \,.
\end{eqnarray}

\begin{figure}
\begin{center}
\includegraphics[width=.8\textwidth]{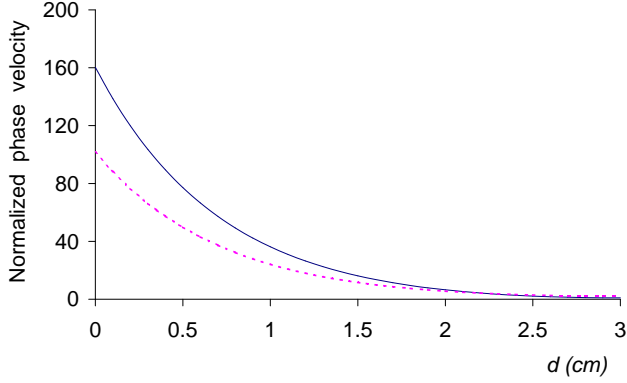}
\end{center}
\caption{Normalised velocities $v_{TE}/c$ (solid line) 
and $v_{TM}/c$ (dotted line), as a function of $d$, for the 
same parameters values as in Fig. 3 and for $\theta_0 =45^\circ$.}
\end{figure}

\section{Conclusions}

On the basis of the vectorial treatment we are now able to compare
the TE and TM cases. In particular we can conclude that:

\begin{description}
\item 1 - The phase shift $\Delta\Phi$  of the transmitted
TM beam (Eq. (\ref{dfi}) with $\Phi^{\prime\prime}$ instead
of $\Phi^\prime$)   
at $z = d$ with respect to the incident beam at $z = 0$
is different from that of the TE case. 
\item 2 - The transmission coefficient $T$  for the TM
case is different in amplitude (Fig. 3a) and phase (Fig. 3b) with
respect to that of the TE case. Consequently, an
incident Bessel beam formed by a TE component and by a
TM component gives rise to a transmitted Bessel beam with a
different polarisation.

We note that the 
amplitudes are slowly varying functions of $\theta_0$,  
for both TE and TM cases. On the contrary, the phase in the case of TE differs 
greatly with respect to the TM case, and both of them greatly vary with $\theta_0$.
\item 3 - The phase velocity $v_{TM}$  and the wavelength 
$\lambda_{TM}$  inside the layer in the TM case are 
different from those of the TE case,

\begin{eqnarray}  
\lambda_{TM}&=& \frac{n\lambda_0}{{n^\prime}^2\cos\theta_0}
\frac{\cosh^2[k_0n^\prime\Gamma(d-z)]}{\cos^2[\eta_{TM}(z)]}\nonumber \\
v_{TM}&=& \frac{n c}{{n^\prime}^2\cos\theta_0}
\frac{\cosh^2[k_0n^\prime\Gamma(d-z)]}{\cos^2[\eta_{TM}(z)]}
\end{eqnarray}
where $\eta_{TM}(z)$ is such that
$\tan[\eta_{TM}(z)] = -\tanh [k_0n^\prime\Gamma (d-z)]\tan\Phi^{\prime\prime}$.
Figure 4 shows the normalised phase velocities $v_{TM}/c$ (which is equal to $\lambda_{TM}/\lambda_0$) and
$v_{TE}/c$ (equal to $\lambda_{TE}/\lambda_0$).
We note that, immediately after the first boundary the motion is extremely fast 
since the effect of the second boundary, which originates the regressive (or ``anti-evanescent'') beam, is negligible. We wish to recall that, in the absence of the second boundary, the phase velocity is infinite.  
\end{description}

\end{document}